# The Effects of Boron Doping on the Bulk and Surface Acoustic Phonons in Single-Crystal Diamond

Erick Guzman,[1] Fariborz Kargar,[1*] Frank Angeles,[2] Reza Vatan Meidanshahi,[3] Timothy A. Grotjohn,[4] Aaron Hardy,[5] Matthias Muehle,[5] Richard B. Wilson,[2] Stephen Goodnik,[3] and Alexander A. Balandin[1*]

[1]Nano-Device Laboratory and Phonon Optimized Engineered Materials Center, Department of Electrical and Computer Engineering, University of California, Riverside, California 92521 USA

[2]Department of Mechanical Engineering and Materials Science and Engineering Program, University of California, Riverside, California 92521 USA

[3]School of Electrical, Computer and Energy Engineering, Arizona State University, Tempe, Arizona 85281 USA

[4]Department of Electrical and Computer Engineering, Michigan State University, East Lansing, Michigan 48824 USA

[5]Fraunhofer USA Center Midwest, East Lansing, Michigan 48824 USA

* Corresponding authors: fkargar@ece.ucr.edu ; balandin@ece.ucr.edu ; web-site: http://balandingroup.ucr.edu/

**ABSTRACT**

We report the results of the investigation of bulk and surface acoustic phonons in the undoped and boron-doped single-crystal diamond films using the Brillouin-Mandelstam light scattering spectroscopy. The evolution of the optical phonons in the same set of samples was monitored with Raman spectroscopy. It was found that the frequency and the group velocity of acoustic phonons decrease non-monotonically with the increasing boron doping concentration, revealing pronounced *phonon softening*. The change in the velocity of the shear horizontal and the high-frequency pseudo-longitudinal acoustic phonons in the degenerately doped diamond, as compared to the undoped diamond, was as large as ~ 15% and ~12%, respectively. As a result of boron doping, the velocity of the bulk longitudinal and transverse acoustic phonons decreased correspondingly. The frequency of the optical phonons was unaffected at low boron concentration but experienced a strong decrease at the high doping level. The density-functional-theory calculations of the phonon band structure for the pristine and highly-doped sample confirm the phonon softening as a result of boron doping in diamond. The obtained results have important implications for thermal transport in heavily doped diamond, which is a promising material for ultra-wide-band-gap electronics.

**INTRODUCTION**

Recent years witnessed a rapid growth of interest in ultra-wide bandgap (UWBG) semiconductors for applications in power electronics.[1–4] The materials, which belong to the UWBG group, include semiconductors such as AlN and diamond, with an electronic bandgap ranging from 3 eV to 6 eV.[3] Among UWBGs, diamond attracts the most attention as it holds a record-high current density, thermal conductivity, mechanical stiffness, chemical stability, and the critical electric field.[5–9] Intrinsic diamond is an electrical insulator; it is doped by boron (B) to become a *p*-type semiconductor, suitable for electronic applications.[10] In diamond, the bulk acoustic phonons, *i.e.,* quanta of crystal lattice vibrations, are the main heat carriers. They have high group velocities and long lifetimes.[7–9] These characteristics are responsible for the diamond's high thermal conductivity of ~2200 $Wm^{-1}K^{-1}$ at room temperature (RT) and excellent thermal interface conductance.[1,7–9,11] The frequency and dispersion of acoustic phonons are also related to the elastic and mechanical properties of the material. While boron substitutional doping improves the electrical conductivity of diamond, it adversely affects its phonon heat conduction characteristics.[12–14] Boron atoms act as point defects, scattering acoustic phonons, shortening their lifetime, and thus reducing the thermal conductivity of the material.[12–15] Although the phonons in diamond have been investigated extensively, the data on the surface and bulk acoustic phonons in doped diamond are scarce and rather controversial. An important open question is "Do boron atoms only act as the scattering centers for acoustic phonons, which retain the frequency and velocity of intrinsic diamond, or do the dopant atoms alter the phonon characteristics of the material themselves?" The effect of doping on the surface acoustic phonons in diamond has also not been addressed. The properties of the surface phonons are important for understanding the thermal transport across interfaces in the device structures.

Brillouin-Mandelstam light scattering (BMS), also referred to as Brillouin light spectroscopy (BLS), is a nondestructive optical technique that has been used extensively to study acoustic phonons in different types of materials.[16] This technique has been employed to examine the mechanical properties of different types of diamond, *e.g.*, polycrystalline, smooth fine-grained, and CVD-grown single-crystal diamond.[17–20] The prior studies reported the characteristics of bulk longitudinal acoustic (LA) and transverse acoustic (TA) phonons as well as the travelling surface acoustic phonons (SAWs) along the high-symmetry crystallographic directions.[19,20] The elasticity coefficients of diamond were extracted from BMS data.[17–20] A recent detailed study has used the angle-resolved BMS to find the properties of SAWs along different crystallographic directions in diamond [21]. No BMS data on the effect of boron on bulk and surface acoustic phonons in boron-doped diamond have been reported to date. In this work, we used the BMS technique to investigate the bulk and surface acoustic phonons in the low, medium, and highly boron-doped CVD-grown diamond films and compared the results to that of the undoped high-pressure high temperature (HPHT) diamond. The evolution of optical phonons in the same samples was monitored with Raman spectroscopy. It was found that the frequency and group velocity of acoustic phonons decrease non-monotonically with the increasing boron doping concentration, revealing pronounced *phonon softening*. The phonon modification with the introduction of dopant atoms appears to be stronger than previously believed. The observed changes in the characteristics of acoustic phonons in diamond because of doping have important broad implications for heat conduction in UWBG materials and for thermal management of UWBG-based electronic devices.[22]

**RESULTS OF MEASUREMENTS**

The single-crystal diamond films for this study have been grown by the chemical vapor deposition (CVD) on the HPHT diamond substrate synthesized via the high-pressure high-temperature method. The details of the growth and boron doping procedures have been reported by some of us elsewhere.[23,24] The quality of the CVD diamond and characteristics of optical phonons were assessed with Raman spectroscopy. The samples with the boron doping concentrations of $10^{16}$ cm$^{-3}$, $10^{17}$ cm$^{-3}$, and $3\times10^{20}$ cm$^{-3}$ were referred to as the low, medium, and highly doped diamond, respectively. Systematic Raman measurements were conducted using a 633-nm wavelength excitation laser in a conventional backscattering configuration. The laser power was kept low at ~60 µW all the time to eliminate any possible laser-induced heating effects. In Figure 1 (a), we present typical Raman spectra for all types of diamond films and the reference HPHT diamond. Comparing the data, one finds additional Raman features appearing as broad peaks in the range of 300 cm$^{-1}$ to ~1300 cm$^{-1}$ in the sample with the highest boron doping concentration. These peaks are activated in diamond as a result of the high boron inclusion resulting in some structural disorder.[25,26] The broad bands near 1200 cm$^{-1}$ and 550 cm$^{-1}$ are generally attributed to the forbidden Raman peaks corresponding to the maximum phonon density of states and the maximum density of states of the acoustic phonons in diamond, respectively.[27] More detailed discussions of these peaks are provided in Ref. [25,27]. In all spectra, the intense peak at ~1332 cm$^{-1}$ originates from the Brillouin zone (BZ) center optical phonons of diamond, referred here as the zone-center phonons (ZCP).[25,27]

Figure 1 (b) shows the details of the ZCP Raman signatures. The intensity of the ZCP peak is lower in the highly doped diamond compared to other samples. The peak is symmetric for all types of samples, and we were able to fit the experimental data accurately using a single Lorentzian function. Several prior Raman studies have reported that boron doping induces an asymmetry in the ZCP peak, which changes its shape from a Lorentzian-like to a Fano-like peak.[25,27] The latter is explained by an increased contribution of scattering by free charge carriers in the highly doped samples.[27] We also observed such evolution of Raman spectra collected under 488-nm and 325-nm wavelength laser excitation (see Supplementary Materials). Since no changes in the symmetry of the ZCP peak were seen in Raman spectra accumulated under the 633-nm laser excitation, we concluded that the behavior of the ZCP peak depends strongly on the wavelength of the excitation laser.

In Figure 1 (c), we show the spectral position and the full-width-at-half-maximum (FWHM) of this Raman peak determined using the Lorentzian fittings. The exact data are also provided in the table format in the Supplementary Materials. The spectral position and the FWHM of the optical phonons remain unchanged for the low and medium boron-doped samples compared to the undoped HPHT reference sample. In the highly doped diamond, however, the ZCP peak red-shifts by ~2 cm$^{-1}$ exhibiting the *optical phonon softening* behavior. Moreover, its FWHM, which inversely relates to the phonon lifetime,[28] increases by a factor of ~3 compared to the undoped HPHT and low and medium doped samples. This is consistent with the observation that structural disorder can also make FWHM wider.[29] If the structural disorder has a length-scale *L*, phonons with the wavelength $q\sim 1/L$ will contribute to the scattering and, as a result, broaden the energy spectrum and lower the average energy of the first-order Raman peak.[30] Possible implications of the optical phonon softening on thermal transport in diamond are discussed below.

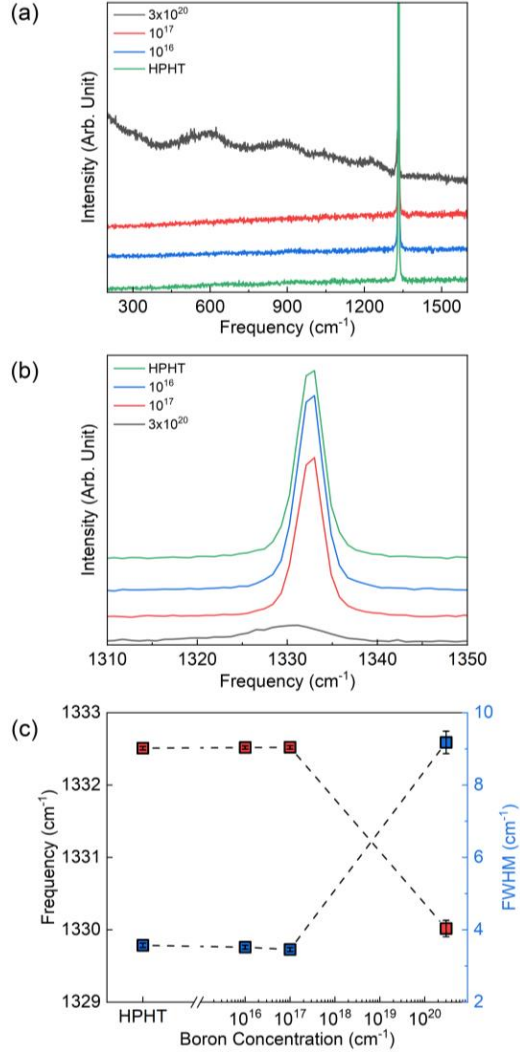

**Figure 1:** (a) Raman spectra of the boron-dope diamond samples and a reference undoped HPHT diamond recorded under a 633-nm excitation laser. (b) Raman spectra for the same samples in the vicinity of the diamond's zone-center optical phonon peak at ~1332 cm$^{-1}$. (c) Spectral position (red squares) and FWHM (blue squares) of the zone-center peak as a function of the boron concentration. The spectral position of the peak decreases by 2 cm$^{-1}$ while its FWHM increases significantly for the highly doped diamond.

We now turn to the main element of this work – BMS investigation of the boron-doped diamond films. The BMS measurements were conducted in the backscattering configuration using a 532 nm excitation laser at a fixed light incident angle $\theta = 20°$. The incident light was *p*-polarized; no specific polarization selection was used for the collection of the scattered light. The details of our BMS procedures are provided in the Methods and prior reports for other material systems.[16,31,32]

The top surfaces of the samples were diamond's (001) crystallographic plane with an off-cut plane angle of ~3°. This small off-cut angle had negligible effect on the light scattering and did not affect the interpretation of BMS data. Figure 2 shows the results of the BMS measurements of the three boron-doped diamond samples and the HPHT diamond substrate in the frequency range of 25 GHz to 250 GHz. Two sharp peaks on each side of the spectra are associated with the Stokes and anti-Stokes scattering processes by the longitudinal acoustic (LA) and the transverse acoustic (TA) bulk phonons. The spectral position of the observed peaks, $f$, was determined accurately by fitting the experimental with individual Lorentzian functions. The details of the fitting procedures are provided in the Supplementary Materials. The phonon wave-vector of these bulk modes is, $q_B = 4\pi n/\lambda$ where $\lambda$ is the wavelength of the excitation laser and $n$ is the refractive index of the medium.[16,33] Table 1 summarizes the peak frequency, FWHM, and the relative intensity of LA to TA modes for all diamond samples.

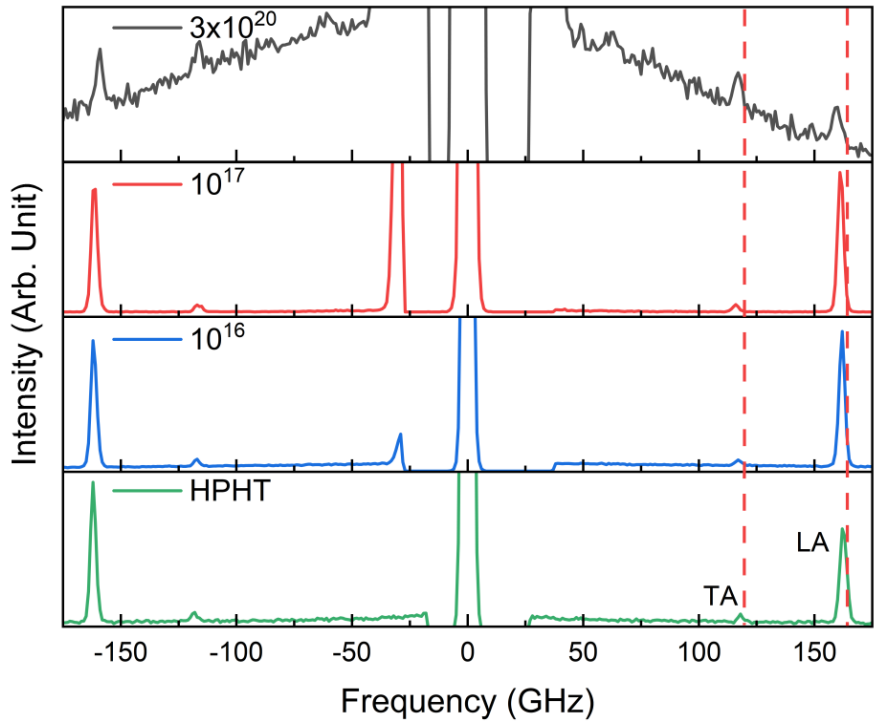

**Figure 2:** Brillouin light scattering spectra of the boron-doped diamond samples in the backscattering geometry performed at a 20° incidence angle. The dashed lines are guides to the eye to illustrate the decrease in the acoustic phonon frequency as the doping level increases. The peaks labeled as LA and TA correspond to the longitudinal acoustic and the transverse acoustic bulk phonons.

Table 1: Spectral position, FWHM, and relative intensity of BMS peaks in three diamond samples

| Boron Doping (cm$^{-3}$) | $f_{TA}$ (GHz) | FWHM$_{TA}$ (GHz) | $f_{LA}$ (GHz) | FWHM$_{LA}$ (GHz) | $I_{LA}/I_{TA}$ |
|---|---|---|---|---|---|
| HPHT | 117.9 | 2.3 | 162.4 | 2.7 | 16.7 |
| 10$^{16}$ | 117.1 | 4.4 | 161.9 | 2.4 | 24.8 |
| 10$^{17}$ | 115.9 | 3.5 | 161.3 | 2.3 | 20.5 |
| 3×10$^{20}$ | 116.9 | 2.9 | 159.5 | 2.6 | 1.0 |

It follows from the data in Table 1 that the LA and TA phonon modes experience softening with an increase in the boron doping level. The LA phonons have a frequency of 162.4 GHz in the undoped HPHT sample. It decreases to 159.5 GHz in the diamond with the highest boron concentration. The frequency of the TA phonons decreases from 117.9 GHz to 116.9 GHz as the boron concentration increases. The FWHM of the peaks does not reveal a clear trend. The latter could be attributed to large experimental uncertainty in determining FWHM. An intriguing observation is that the relative intensity of the LA phonon peak with respect to the TA phonon, $I_{LA}/I_{TA}$, decreases substantially for the highest doped diamond. In optically isotropic materials and in the backscattering BMS configuration, the spectral power scattered by TA phonons falls to zero and therefore, the spectrum is dominated by the LA peak.[33] In our results, this is the case for the low and medium-doped diamond samples. However, in high boron-doped diamond, the scattering intensity of the LA peaks is suppressed significantly, becoming even weaker than the associated TA peak. Given that the $I_{LA}/I_{TA}$ ratio in the BMS strongly depends on the boron concentration, one can use this parameter to determine the local boron doping with a spatial resolution of 25 μm and 1 μm using the regular- and micro-BMS systems, respectively.

Knowing the frequency of the phonon modes, $f$, and the probed phonon wave-vector, $q_B$, one can also obtain the phase velocity, $v_P$, of the phonons as $v_p = 2\pi f/q_B$. Note that since the dispersion of acoustic phonons close to the Brillouin zone (BZ) center is linear, *i.e.*, $\omega = qv_p$, the phase velocity, $v_p = \omega/q$, and the group velocity, $v_g = \partial\omega/\partial q$, of the fundamental LA and TA acoustic phonons are essentially the same, *i.e.*, $v_p = v_g$. In our experiments, the direction of the examined phonon wave-vector lies close to the [001] real-space crystallographic direction with a deviation angle of $\theta^*$ from [001] direction. The deviation angle can be calculated using the Snell's law, as $\sin(\theta^*) = \sin(\theta)/n$, in which $n$ is the diamond's index of refraction and $\theta = 20°$ is the incident angle fixed for all measurements. The characteristics of acoustic phonons depend on the crystallographic direction and thus, possible changes due to $\theta^*$ should be considered. We measured the change in the refractive index of the boron-doped diamond samples due to the boron dopants in order to account for this effect (see the Methods section). For the low and medium doped samples, the refractive index, measured at 515 nm wavelength, was 2.43, whereas for the highest doped sample it reduced slightly to 2.41. The measured values of $n$, and the respective deviation angle and the calculated group velocity of TA and LA phonons are summarized in Table 2. The calculated values for the reference HPHT sample are in good agreement with the reported data in literature.[18,21] Note that the calculated values of $v_g$ may inherit some error due to the measurement of $n$ at a wavelength slightly different from that used in BMS experiments. We expect a similar variation in $n$ at the laser excitation wavelength used in BMS. From these data, one can see that the group velocity of TA and LA phonons decrease with the boron incorporation. To further support our conclusions and reduce the uncertainty due to measurements of $n, \theta^*$ and Lorentzian fitting, which are always present in experiments with bulk phonons, we examined closely the characteristics of the surface acoustic phonons in the undoped and doped diamond samples.

Table 2: The index of refraction and group velocity of fundamental phonon modes

| Boron Doping ($cm^{-3}$) | $n$ @ $515\ nm$ | $\theta^*$ (degrees) | $v_{g,TA}$ ($ms^{-1}$) | $v_{g,LA}$ ($ms^{-1}$) |
|---|---|---|---|---|
| HPHT | 2.3812* | 8.26 | 13170.42 | 18141.44 |
| $10^{16}$ | 2.43 | 8.09 | 12844.56 | 17713.95 |
| $10^{17}$ | 2.43 | 8.09 | 12678.35 | 17650.52 |
| $3\times10^{20}$ | 2.41 | 8.17 | 12683.75 | 17651.62 |

* Ref. [34]

In BMS experiments, light can also be scattered by the propagating surface ripples caused by the displacement fields of surface phonons or reflected bulk phonons from the interfaces.[16,33,35] In this scattering mechanism, the phonon wavevector, defined as $q_{||} = 4\pi \sin(\theta)/\lambda$, lies parallel to the surface of the sample. This wavevector only depends on the light incident angle, $\theta$, and the wavelength of the excitation laser in vacuum, $\lambda$. It is important to note that in this case, the probing phonon wavevector is no longer a function of the refractive index, $n$. For this reason, all the uncertainties, associated with the calculation of the velocities for bulk phonons and deviation angles from the high symmetry [001] crystallographic direction, are eliminated. Figure 3 presents the results of the surface Brillouin scattering measurements performed at a constant incident angle of $\theta = 75°$. The calculated in-plane phonon wavevector is $q_{||} = 0.0228 \text{ nm}^{-1}$. Two pronounced peaks are attributed to the shear-horizontal surface wave (SHW) and high-frequency pseudo-longitudinal wave (HFPLW).[17,21] The important observation is that the frequencies of these peaks decrease with increasing boron doping concentration in diamond. The latter provides solid evidence for the phonon softening in the acoustic polarization branches.

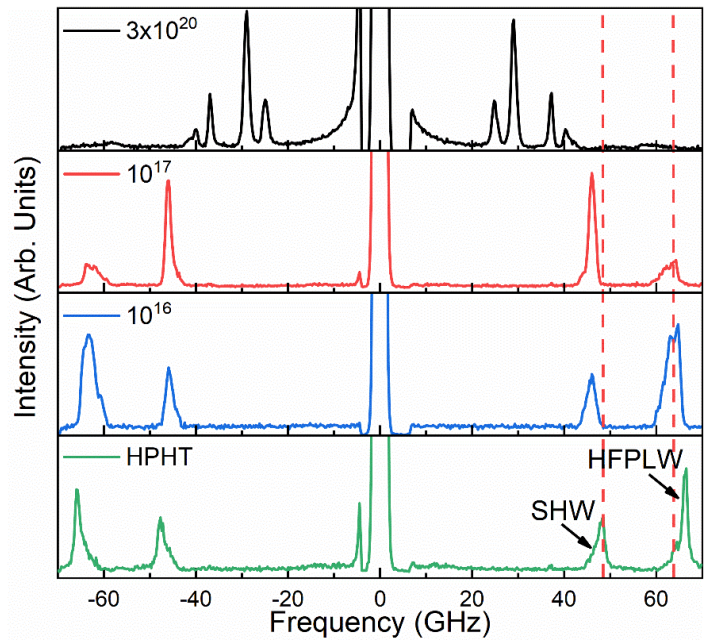

**Figure 3:** Brillouin light scattering spectra of the boron-doped diamond samples and a reference HPHT diamond substrate accumulated at $\theta = 75°$. The shear-horizontal surface wave and high-frequency pseudo-longitudinal wave are labeled as SHW and HFPLW, respectfully. The frequency of types of phonons decreases as the boron concentration increases.

The frequency, $f$, and the phase velocity of the SAW phonons, $v_p = 2\pi f/q_{||}$, are presented as the functions of the boron concentration in Figure 4. One can see that doping the diamond with boron at low concentration of $10^{16}$ cm$^{-3}$, makes the phase velocity of both phonon branches to decrease as compared to that of the undoped reference HPHT diamond. Additional doping of the diamond sample up to $10^{17}$ cm$^{-3}$ does not strongly affect the phonon velocity. However, for the heavily doped diamond, with a boron concentration of $3\times10^{20}$ cm$^{-3}$, another strong decrease in the phase velocity is observed. This non-monotonic trend is similar to what has been reported previously by some of us for the phonon softening in the doped alumina samples.[36] In the alumina, the low mass and small radius aluminum (Al) atoms were substituted with the heavy and large radius neodymium (Nd) atoms.[36] In the boron-doped diamond, the situation is different. On the atomic scale, boron is slightly lighter than carbon, *i.e.*, $m_B/m_C \sim 0.91$, and therefore, the variation in mass cannot explain the observed phonon softening. However, boron's radius is slightly bigger than that of carbon, $R_B/R_C \sim 1.3$. Therefore, a possible mechanism for the phonon softening is the lattice distortion induced by the boron atoms. This distortion, especially in the heavily doped sample, causes an increase in the "effective" crystal lattice parameter and the atomic plane separation. This agrees well with the redshift observed in the Raman ZCP peak of diamond (see Figure 1 (b)). An independent study confirms that the lattice constant of diamond slightly increases with the boron dopant concentration and starts to vary at a higher rate as the concentration surpasses $\sim 2.7 \times 10^{20}$ cm$^{-3}$.[37] Comparing the phonon phase velocity of the highly-doped sample with that of the undoped HPHT diamond, one can see that phonon velocity of the SHW and HFPLW polarization branches decreases by more than ~15% and ~11.7%, respectively.

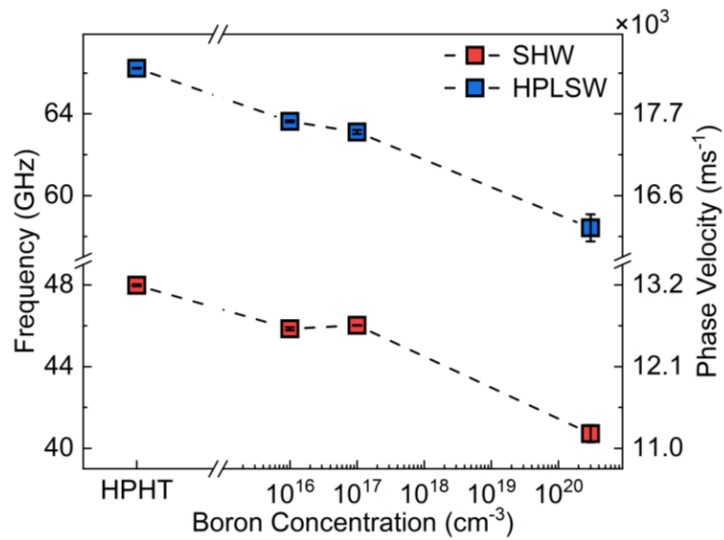

**Figure 4:** Frequency and velocity of surface phonons in diamond as a function of the boron doping concentration. The data for the SHW and HPLSW phonon polarization branches are shown at $q_{||} = 0.0228$ nm$^{-1}$.

In diamond, the phase velocity of the HFPLW is negligibly smaller than that of the bulk LA phonons traveling in the <001> direction.[18] The observed reduction in $v_P$ of the HFPLW with the boron doping concentration points out that the phase velocity, and correspondingly the group velocity of the respective LA phonon polarization branch experience similar softening. Such a reduction has important implications for the thermal and mechanical properties of boron-doped diamond. The phase velocity of SHW and HFPLW in diamond can be estimated from the equations $v_{SHW} = (c_{66}/\rho)^{1/2}$ and $v_{HFPLW} = (c_{11}/\rho)^{1/2}$, respectively, where $\rho$ is the mass density.[18] Note that the phase velocities of SHW and HFPLW phonons depend only on one elasticity constant. Therefore, we can directly relate the elasticity parameters $c_{66}$ and $c_{11}$ to the measured phase velocities of these surface acoustic waves. Assuming a constant mass density for the boron-doped diamond samples, one can infer that the elastic constants decrease with boron addition, in line with the previously reported values obtained by other techniques.[38,39]

**RESULTS OF MODELING**

We performed phonon band structure calculations using the density functional theory (DFT) implemented in Quantum Espresso 7.0 package[40] to investigate the effect of B-doping on the phonon energy dispersion in diamond. The details of the calculations can be found in METHODS section. We used a cube supercell containing 8 atoms in the DFT calculations. For B-doped diamond, one of the C atoms was replaced by B, equivalent to the ~2×10$^{22}$ cm$^{-3}$ doping concentration (Figure 5 (a-b)). Note that we intentionally assumed higher boron concentrations in the simulations to keep the size of the supercell and the cost of calculations reasonable. In order to validate our modeling approach, the phonon dispersion of pure diamond was calculated using an FCC unit cell containing 2 carbon atoms with the experimental lattice constant of 3.57 Å. The computational results were compared with the experimental data from Ref. [41]. The data presented in the Supplemental Materials indicates an excellent agreement between the calculated and experimental phonon frequencies.

After relaxation, the optimized lattice constants were obtained to be 3.54 Å and 3.60 Å for pure and B-doped supercell, respectively. The obtained lattice constant of pure diamond agrees well with the measured lattice constant of 3.57 Å.[37] Figure 5 (c) shows the calculated phonon dispersion of pure diamond in comparison with that of the B-doped diamond. Since we used the supercell and thus, smaller Brillouin zone, all phonon bands are folded back in the reciprocal space. In order to make the phonon dispersion easier to understand, we unfolded the supercell dispersion using the Phonon Unfolding package.[42] As a result, the phonon dispersion of pure diamond is completely reproduced and the excess phonon modes at high symmetry points for both pure and B-doped diamond have been almost completely removed. As seen in Figure 5 (c), both the LA and TA phonon modes undergo softening as a result of the doping. The calculated phonon velocities in pristine diamond for LA and TA modes along the $\Gamma - X$ direction and close to the BZ center are 17494 ms$^{-1}$ and 12388 ms$^{-1}$, respectively. The obtained values agree well with our experimental measurements for the undoped HPHT diamond. In the B-doped diamond, the velocity for the LA and TA phonons decreases substantially. The change in the group velocity of LA polarization branch is more pronounced compared to the TA mode, which is in a qualitative agreement with our experimental observation. Note that since we considered two-orders of magnitude higher boron concentration in our simulations, we do not make a direct quantitative comparison of the velocities.

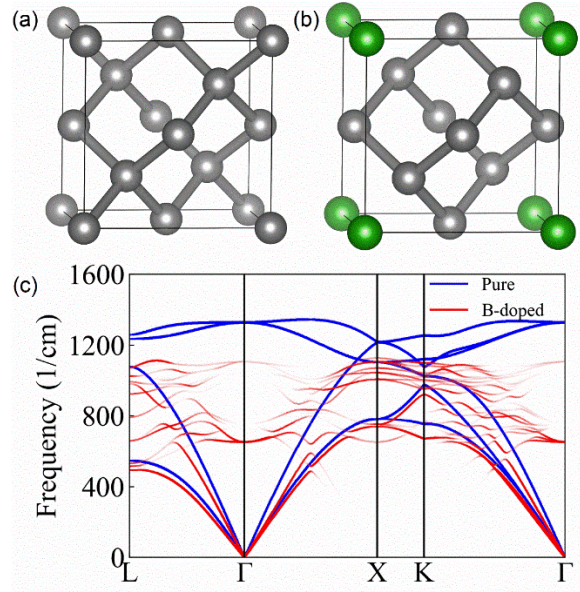

**Figure 5:** (a-b) The supercell structure of the pristine and the B-doped diamond. The grey and green spheres represent C and B atoms, respectively. (c) The calculated phonon band structure of the pristine (blue) and B-doped (red) diamond. The softening of all acoustic phonon modes as a result of boron doping is clearly observed. Note that we assumed substantially higher concentration of B atoms in order to keep the supercell reasonably small in the calculations.

**DISCUSSION**

The phonon softening has interesting and important implications for thermal transport and related properties. Conventional theories of thermal conductivity of semiconductor and insulating materials assumes that doping does not modify the group velocity, $v_g$, of the acoustic phonons.[43–45] The dopants act as an extra point defect scattering centers for the acoustic phonons, which retain their properties the same as in the intrinsic material. However, if the phonon group velocity changes as a result of the doping, one needs to take it into account in the calculations.[46] In the kinetic theory, the phonon thermal conductivity can be expressed as $K = (1/3)Cv_g\Lambda = (1/3)Cv_g^2\tau$, where $C$ is the volumetric heat capacity, $v_g$ is the average phonon group velocity, $\Lambda$ is the phonon "grey" mean-free path, and $\tau$ is the combined phonon relaxation time, *i.e.* lifetime. The phonon relaxation time is defined by scattering rate in different types of processes; it can be expressed as $\tau^{-1} = \tau_U^{-1} + \tau_d^{-1} + \tau_{e-ph}^{-1}$. Here, $\tau_U^{-1}$, $\tau_d^{-1}$, and $\tau_{e-ph}^{-1}$ are the phonon – phonon Umklapp scattering rate due to the crystal anharmonicity, phonon – defect scattering rate, and the electron-phonon scattering rate, respectively. The scattering rates are the function of both phonon frequency, $\omega$, and phonon group velocity, $v_g$. The phonon – defect scattering, which includes scattering on dopant atoms, is the dominate mechanism at low temperatures.[47] Given the high Debye temperature of diamond ($\theta_D \sim 1870$ K), the phonon – defect scattering makes a significant contribution even at room temperature (RT). The phonon softening alters the thermal conductivity, $K$, not only *via* the combined relaxation time, $\tau$, but also *via* the volumetric heat capacity, $C$, which is proportional to $v_g^{-3}$. The dependence on the phonon group velocity is due to the phonon density of states. The previously reported data on the heat capacity of boron-doped single-crystal HPHT diamond support our arguments.[48] In this study, the doped diamond samples with higher boron concentration reveled higher heat capacity. The authors have interpreted their observation as a possible inclusion of metallic particles during the growth process and thus, the dominance of electron heat capacity at low temperatures.[48] However, their data can be also explained by the lower phonon group velocity. One should remember that in the bulk crystals the phonon group velocity is the same as the phonon phase velocity, and the observed reduction in the phonon velocity is equivalent for both.

The strongest effect from the phonon softening on thermal conductivity near RT is expected via the changes in the phonon – defect scattering rate. The average phonon group velocity for all three acoustic polarization branches is given by the expression:[49]

$$v_g = 3\{(1/v_{T,1}) + (1/v_{T,2}) + (1/v_L)\}^{-1}. \tag{1}$$

Here $v_{T,1}$ and $v_{T,2}$ are the phonon group velocities for two TA phonon polarization branches, and $v_L$ is the LA phonon group velocity. Given that in diamond transverse acoustic phonons along <001> direction are degenerate, our experimental data show that all velocities that enter Eq. (1) experience a reduction in their values. The phonon scattering rate on point defects, like dopant atoms, is given as:[50,51]

$$\tau_P^{-1} = (V_0 \Gamma \omega^4 / 4\pi v_g^3) \ . \tag{2}$$

Here $V_0$ is the volume per atom, and $\Gamma$ is the scattering parameter, which depends linearly on the defect concentration. $\Gamma$ is a measure of the point defect scattering strength.[52,53] If the properties of phonons had not change, the increase in this scattering rate would be only due to the increase in the concentration of dopant atoms. As a result of the reduction in the phonon velocity, $v_g$, the effect of the doping becomes stronger and likely dependent on crystallographic direction.

It is also illustrative to consider the effect of the acoustic phonon softening for dislocation scattering. At high density of doping, the dopant atoms may form clusters of atoms that act similar to dislocations lines. The phonon scattering rate on dislocation lines is give as:[49,54]

$$\tau_E^{-1} = (2^{1.5}/3^{3.5}) \eta N_D^E b_E^2 \gamma^2 \omega \{(1/2) + (1/24)[(1-2v)/(1-v)]^2 \left[1 + \sqrt{2}(v_L/v_T)^2\right]^2\} \tag{3}$$

Here, $N_D^E$ and $b_E$ are the density and magnitude of Burgers vectors for the edge dislocations, respectively, $\eta$ accounts for the orientation of the dislocation lines with respect to the direction of temperature gradient. $\gamma$ and $v$ are the Grüneisen anharmonicity parameter and Poisson ratio, respectively. One can see from Eq (3) that even a relative reduction in the phonon velocity for the LA and TA phonons can play a role since the scattering rate depends on $\sim (v_L/v_T)^4$. The discussed dependencies result in intricate effects of phonon softening on heat conduction in diamond. They can lead to stronger reduction in the thermal conductivity that the predictions of simple theory that does not account for the changes in the phonon velocity or more pronounced anisotropy in thermal conductivity if the dopant atoms have some preferential arrangements.

The effects of phonon softening on thermal conductivity are not limited to acoustic phonons alone. Indeed, the contribution of the optical phonons to the phonon thermal conductivity of diamond is negligible owing to their high frequency and small group velocity.[55,56] However, optical phonons provide scattering channels for the acoustic phonon branches.[9] The broadening of the ZCP peak in the highly doped diamond sample indicates a shortened phonon lifetime due to the increased contribution of boron atoms to the phonon scattering processes via point defects and electron-phonon processes. One can estimate the decay rate of the longitudinal optical (LO) phonons at the Brillouin zone-center ($q = 0$) using the Klemens' formula $\tau^{-1} \sim \delta\omega$, where $\tau$ is the phonon lifetime and $\delta\omega$ is the FWHM of the Raman ZCP peak.[56,57] The estimated values of $\tau$ for the undoped HPHT and the low and medium doped diamond samples is ~11 ps, which is close to the measured lifetime of phonons in single-crystal diamond using other techniques.[58] Our data shows that this value decreases to ~3.7 ps for the highly doped sample, significantly lower than that of the HPHT and other lightly doped diamond samples. This reduction can be correlated well with the measured reduction in the thermal conductivity.

**CONCLUSIONS**

We investigated the bulk and surface acoustic phonons in the boron-doped single-crystal diamond films using the Brillouin-Mandelstam light scattering spectroscopy. It was found that the frequency and the group velocity of acoustic phonons decrease non-monotonically with the increasing boron doping concentration, revealing pronounced *phonon softening*. As a result of boron doping, the velocity of the bulk longitudinal and transverse acoustic phonons decreased correspondingly. The frequency of the optical phonons was unaffected at low boron concentration but experienced a strong decrease at the high doping level. We also performed density-functional-theory to calculate the phonon band structure for the pristine and highly doped diamond. The theoretical results qualitatively confirm the phonon softening in both optical and acoustic polarization branches. The strong softening of the acoustic phonons – the main heat carriers in diamond – have important implications for thermal transport in such materials. Our results also demonstrate that the intensity ratio of the LA and TA phonons can be used to monitor the boron concentration in diamond.

**METHODS**

**Boron-doped diamond growth:** The boron doped diamond was grown by microwave plasma-assisted chemical vapor deposition using hydrogen and methane feed-gases with the boron added during the homoepitaxial growth using diborane feed-gas. The substrates used were HPHT seeds that were prepared with an off-cut angle of the growth surface of ~3° from the (001) crystal plane. The boron concentrations were estimated based on similarly grown samples grown with equivalent conditions that were measured by Secondary Ion Mass Spectroscopy (SIMS) at EAG.

**Brillouin – Mandelstam spectroscopy:** BMS experiments were conducted in the conventional backscattering configuration using a 532 nm laser excitation wavelength at several different incident angles. The light source was a solid-state diode-pumped continues-wave laser (Spectra Physics). The laser beam was focused on the sample using a lens with NA=0.34. The scattered light was collected via the same lens and directed to the high-contrast high-resolution 3 + 3 pass tandem Fabry-Perot interferometer (TFP-1, JRS Optical Instruments, Switzerland) and spectrometer. For bulk and surface acoustic phonon measurements, the mirror-spacing of the TFP was adjusted on 0.5 mm and 2 mm, respectively.

**Refractive index measurements:** The refractive index of the diamond thin films was determined by measuring the Fresnel reflectance coefficients for s and p polarized 515 nm light. For the reflectance measurements, we use a linearly polarized laser beam with a polarization 45° from the vertical. We focused the light on the sample using a 10× objective lens. The angle of incidence of the beam with the sample was kept 45°. Light reflected from the sample was collected by another 10× objective lens. Light that reflected from the samples back surface was spatially separated from the primary reflected beam with an aperture. The intensity and polarization of the reflected probe beam was measured by a calibrated power meter and a polarimeter. We used a multilayer optical calculation to analyze the reflectance data. The index of refraction of the diamond film was treated as a fitting parameter until the model predictions matched the experimental observations for the polarization and intensity of the reflected laser beam.

**Density-Functional-Theory (DFT) calculations:** The ultra-soft pseudopotential and the generalized gradient approximation (GGA) of Perdew–Burke–Ernzerhof (PBE)[59] functional were employed in order to describe the core-valence electrons interaction and exchange correlation energy, respectively. The kinetic energy cutoff of 25 Ry was chosen for plane wave basis sets. We use a cube supercell containing 8 atoms in our DFT calculations. For boron doped diamond, one of the C atoms was replace by B which was equivalent to the ~$2\times10^{22}$ cm$^{-3}$ doping concentration. Both supercell size and atomic positions were relaxed using BFGS quasi-newton algorithm within DFT until the force on each atom was smaller than 0.001 Ry/Bohr. In the self-consistent ground state calculations, 8x8x8 Monkhorst–Pack k-points setting was used in the reciprocal space integration. After obtaining the self-consistent ground state, we performed the density functional perturbation theory (DFPT)[60] calculations with a uniform 4x4x4 grid of q-points setting. Then we performed dynamical matrix Fourier transformations to obtain force constants, and calculated the phonon dispersions. In order to benchmark our method of calculations, the phonon dispersion of pure diamond was calculated using an FCC unit cell containing 2 carbon atoms with the experimental lattice constant of 3.57 Å (Figure SX) and were compared with experimental data.[41]

## ASSOCIATED CONTENT

**Supporting Information**

The supporting information is available free of charge on the ACS Publication website at DOI: XXXX

The supporting information includes a detailed description of the Raman and BMS measurements.

■ **AUTHOR INFORMATION**

**Corresponding Author**

\* Email: balandin@ece.ucr.edu (A.B.B.)

\* Email: fkargar@ece.ucr.edu (F.K.)

**ORCID**

Fariborz Kargar: 0000-0003-2192-2023

Alexander A. Balandin: 0000-0002-9944-7894

Timothy A. Grotjohn 0000-0003-3500-538119 | P a g e

**ACKNOWLEDGEMENTS**

The work at UCR, ASU and MSU was supported by ULTRA, an Energy Frontier Research Center (EFRC) funded by the U.S. Department of Energy, Office of Science, Basic Energy Sciences under Award #  DE-SC0021230. The authors thank Prof. Robert Nemanich, ASU for useful discussions.

**CONTRIBUTIONS**

A.A.B. and F.K. conceived the idea, coordinated the project, contributed to experimental data analysis, and led the manuscript preparation; E.G. conducted Raman and BMS experiments and contributed to the data analysis; F.K. supervised the BMS experiments; F.A. conducted the refractive index measurements; T.G., A.H., and M.M. synthesized the HPHT and B-doped diamond samples; R.W. supervised the refractive index measurements and assisted with the data analyses; R.M. conducted numerical simulations; S.G. provided the theoretical calculations and assisted with data analysis. All authors contributed to the manuscript preparation.

# REFERENCES

(1) Malakoutian, M.; Field, D. E.; Hines, N. J.; Pasayat, S.; Graham, S.; Kuball, M.; Chowdhury, S. Record-Low Thermal Boundary Resistance between Diamond and GaN-on-SiC for Enabling Radiofrequency Device Cooling. *ACS Applied Materials and Interfaces* **2021**, *13* (50), 60553–60560. https://doi.org/10.1021/acsami.1c13833.

(2) Surdi, H.; Thornton, T.; Nemanich, R. J.; Goodnick, S. M. Space Charge Limited Corrections to the Power Figure of Merit for Diamond. *Applied Physics Letters* **2022**, *120* (22), 223503. https://doi.org/10.1063/5.0087059.

(3) Ghosh, S.; Surdi, H.; Kargar, F.; Koeck, F. A.; Rumyantsev, S.; Goodnick, S.; Nemanich, R. J.; Balandin, A. A. Excess Noise in High-Current Diamond Diodes. *Applied Physics Letters* **2022**, *120* (6), 062103. https://doi.org/10.1063/5.0083383.

(4) Tsao, J. Y.; Chowdhury, S.; Hollis, M. A.; Jena, D.; Johnson, N. M.; Jones, K. A.; Kaplar, R. J.; Rajan, S.; van de Walle, C. G.; Bellotti, E.; Chua, C. L.; Collazo, R.; Coltrin, M. E.; Cooper, J. A.; Evans, K. R.; Graham, S.; Grotjohn, T. A.; Heller, E. R.; Higashiwaki, M.; Islam, M. S.; Juodawlkis, P. W.; Khan, M. A.; Koehler, A. D.; Leach, J. H.; Mishra, U. K.; Nemanich, R. J.; Pilawa-Podgurski, R. C. N.; Shealy, J. B.; Sitar, Z.; Tadjer, M. J.; Witulski, A. F.; Wraback, M.; Simmons, J. A. Ultrawide-Bandgap Semiconductors: Research Opportunities and Challenges. *Advanced Electronic Materials* **2018**, *4* (1), 1600501. https://doi.org/10.1002/aelm.201600501.

(5) Donato, N.; Rouger, N.; Pernot, J.; Longobardi, G.; Udrea, F. Diamond Power Devices: State of the Art, Modelling, Figures of Merit and Future Perspective. *Journal of Physics D: Applied Physics* **2019**, *53* (9), 093001. https://doi.org/10.1088/1361-6463/AB4EAB.

(6) Surdi, H.; Koeck, F. A. M.; Ahmad, M. F.; Thornton, T. J.; Nemanich, R. J.; Goodnick, S. M. Demonstration and Analysis of Ultrahigh Forward Current Density Diamond Diodes. *IEEE Transactions on Electron Devices* **2022**, *69* (1), 254–261. https://doi.org/10.1109/TED.2021.3125914.

(7) Wei, L.; Kuo, P. K.; Thomas, R. L.; Anthony, T. R.; Banholzer, W. F. Thermal Conductivity of Isotopically Modified Single Crystal Diamond. *Physical Review Letters* **1993**, *70* (24), 3764–3767. https://doi.org/10.1103/PhysRevLett.70.3764.

(8)  Olson, J. R.; Pohl, R. O.; Vandersande, J. W.; Zoltan, A.; Anthony, T. R.; Banholzer, W. F. Thermal Conductivity of Diamond between 170 and 1200 K and the Isotope Effect. *Physical Review B* **1993**, *47* (22), 14850–14856. https://doi.org/10.1103/PhysRevB.47.14850.

(9)  Ward, A.; Broido, D. A.; Stewart, D. A.; Deinzer, G. Ab Initio Theory of the Lattice Thermal Conductivity in Diamond. *Physical Review B - Condensed Matter and Materials Physics* **2009**, *80* (12), 125203. https://doi.org/10.1103/PhysRevB.80.125203.

(10) Kalish, R. Doping of Diamond. *Carbon* **1999**, *37* (5), 781–785. https://doi.org/10.1016/S0008-6223(98)00270-X.

(11) Mandal, S.; Yuan, C.; Massabuau, F.; Pomeroy, J. W.; Cuenca, J.; Bland, H.; Thomas, E.; Wallis, D.; Batten, T.; Morgan, D.; Oliver, R.; Kuball, M.; Williams, O. A. Thick, Adherent Diamond Films on AlN with Low Thermal Barrier Resistance. *ACS Applied Materials and Interfaces* **2019**, *11* (43), 40826–40834. https://doi.org/10.1021/acsami.9b13869.

(12) Prikhodko, D.; Tarelkin, S.; Bormashov, V.; Golovanov, A.; Kuznetsov, M.; Teteruk, D.; Volkov, A.; Buga, S. Thermal Conductivity of Synthetic Boron-Doped Single-Crystal HPHT Diamond from 20 to 400 K. *MRS Communications* **2016**, *6* (2), 71–76. https://doi.org/10.1557/MRC.2016.12.

(13) Prikhodko, D.; Tarelkin, S.; Bormashov, V.; Golovanov, A.; Kuznetsov, M.; Teteruk, D.; Kornilov, N.; Volkov, A.; Buga, A. Low Temperature Thermal Conductivity of Heavily Boron-Doped Synthetic Diamond: Influence of Boron-Related Structure Defects. *Journal of Superhard Materials 2019 41:1* **2019**, *41* (1), 24–31. https://doi.org/10.3103/S1063457619010039.

(14) Williams, G.; Calvo, J. A.; Faili, F.; Dodson, J.; Obeloer, T.; Twitchen, D. J. Thermal Conductivity of Electrically Conductive Highly Boron Doped Diamond and Its Applications at High Frequencies. *Proceedings of the 17th InterSociety Conference on Thermal and Thermomechanical Phenomena in Electronic Systems, ITherm 2018* **2018**, 235–239. https://doi.org/10.1109/ITHERM.2018.8419493.

(15) Ding, M.; Liu, Y.; Lu, X.; Li, Y.; Tang, W. Boron Doped Diamond Films: A Microwave Attenuation Material with High Thermal Conductivity. *Applied Physics Letters* **2019**, *114* (16), 162901. https://doi.org/10.1063/1.5083079.

(16) Kargar, F.; Balandin, A. A. Advances in Brillouin–Mandelstam Light-Scattering Spectroscopy. *Nature Photonics* **2021**, *15* (10), 720–731. https://doi.org/10.1038/s41566-021-00836-5.

(17) Djemia, P.; Tallaire, A.; Achard, J.; Silva, F.; Gicquel, A. Elastic Properties of Single Crystal Diamond Made by CVD. *Diamond and Related Materials* **2007**, *16* (4–7), 962–965. https://doi.org/10.1016/J.DIAMOND.2006.11.029.

(18) Djemia, P.; Dugautier, C.; Chauveau, T.; Dogheche, E.; de Barros, M. I.; Vandenbulcke, L. Mechanical Properties of Diamond Films: A Comparative Study of Polycrystalline and Smooth Fine-Grained Diamonds by Brillouin Light Scattering. *Journal of Applied Physics* **2001**, *90* (8), 3771–3779. https://doi.org/10.1063/1.1402667.

(19) Grimsditch, M. H.; Ramdas, A. K. Brillouin Scattering in Diamond. *Physical Review B* **1975**, *11* (8), 3139–3148. https://doi.org/10.1103/PhysRevB.11.3139.

(20) Jiang, X.; Harzer, J. v.; Hillebrands, B.; Wild, Ch.; Koidl, P. Brillouin Light Scattering on Chemical-Vapor-Deposited Polycrystalline Diamond: Evaluation of the Elastic Moduli. *Applied Physics Letters* **1991**, *59* (9), 1055–1057. https://doi.org/10.1063/1.106343.

(21) Xie, Y. R.; Ren, S. L.; Gao, Y. F.; Liu, X. L.; Tan, P. H.; Zhang, J. Measuring Bulk and Surface Acoustic Modes in Diamond by Angle-Resolved Brillouin Spectroscopy. *Science China: Physics, Mechanics and Astronomy* **2021**, *64* (8), 287311. https://doi.org/10.1007/s11433-020-1710-6.

(22) Turin, V. O.; Balandin, A. A. Electrothermal Simulation of the Self-Heating Effects in GaN-Based Field-Effect Transistors. *Journal of Applied Physics* **2006**, *100* (5), 054501. https://doi.org/10.1063/1.2336299.

(23) Ramamurti, R.; Becker, M.; Schuelke, T.; Grotjohn, T. A.; Reinhard, D. K.; Asmussen, J. Deposition of Thick Boron-Doped Homoepitaxial Single Crystal Diamond by Microwave Plasma Chemical Vapor Deposition. *Diamond and Related Materials* **2009**, *18* (5–8), 704–706. https://doi.org/10.1016/J.DIAMOND.2009.01.031.

(24) Demlow, S. N.; Rechenberg, R.; Grotjohn, T. The Effect of Substrate Temperature and Growth Rate on the Doping Efficiency of Single Crystal Boron Doped Diamond. *Diamond and Related Materials* **2014**, *49*, 19–24. https://doi.org/10.1016/J.DIAMOND.2014.06.006.

(25) Prawer, S.; Nemanich, R. J. Raman Spectroscopy of Diamond and Doped Diamond. *Philosophical Transactions of the Royal Society of London. Series A: Mathematical, Physical and Engineering Sciences* **2004**, *362* (1824), 2537–2565. https://doi.org/10.1098/RSTA.2004.1451.

(26) Ferrari, A. C.; Robertson, J. Resonant Raman Spectroscopy of Disordered, Amorphous, and Diamondlike Carbon. *Physical Review B - Condensed Matter and Materials Physics* **2001**, *64* (7), 075414. https://doi.org/10.1103/PhysRevB.64.075414.

(27) Mortet, V.; Živcová, Z. V.; Taylor, A.; Davydová, M.; Frank, O.; Hubík, P.; Lorincik, J.; Aleshin, M. Determination of Atomic Boron Concentration in Heavily Boron-Doped Diamond by Raman Spectroscopy. *Diamond and Related Materials* **2019**, *93*, 54–58. https://doi.org/10.1016/j.diamond.2019.01.028.

(28) Menéndez, J.; Cardona, M. Temperature Dependence of the First-Order Raman Scattering by Phonons in Si, Ge, and -Sn: Anharmonic Effects. *Physical Review B* **1984**, *29* (4), 2051–2059. https://doi.org/10.1103/PhysRevB.29.2051.

(29) Balkas, C.; Shin, H.; Davis, R.; Nemanich, R. Raman Analysis of Phonon Lifetimes in Aln and Gan of Wurtzite Structure. *Physical Review B N.* **1999**, *59* (20), 12977–12982. https://doi.org/10.1103/PhysRevB.59.12977.

(30) Piscanec, S.; Cantoro, M.; Ferrari, A. C.; Zapien, J. A.; Lifshitz, Y.; Lee, S. T.; Hofmann, S.; Robertson, J. Raman Spectroscopy of Silicon Nanowires. *Physical Review B - Condensed Matter and Materials Physics* **2003**, *68* (24), 241312–242003. https://doi.org/10.1103/PhysRevB.68.241312.

(31) Kargar, F.; Debnath, B.; Kakko, J.-P.; Saÿnätjoki, A.; Lipsanen, H.; Nika, D. L.; Lake, R. K.; Balandin, A. A. Direct Observation of Confined Acoustic Phonon Polarization Branches in Free-Standing Semiconductor Nanowires. *Nature Communications* **2016**, *7*, 13400. https://doi.org/10.1038/ncomms13400.

(32) Huang, C. Y. T.; Kargar, F.; Debnath, T.; Debnath, B.; Valentin, M. D.; Synowicki, R.; Schoeche, S.; Lake, R. K.; Balandin, A. A. Phononic and Photonic Properties of Shape-Engineered Silicon Nanoscale Pillar Arrays. *Nanotechnology* **2020**, *31* (30), 30LT01. https://doi.org/10.1088/1361-6528/ab85ee.

(33) Sandercock, J. R. Trends in Brillouin Scattering: Studies of Opaque Materials, Supported Films, and Central Modes. In *Light Scattering in Solids III. Topics in Applied Physics*;

Cardona, M., Güntherodt, G., Eds.; Springer: Berlin, Heidelberg, 1982; Vol. 51, pp 173–206. https://doi.org/10.1007/3540115137_6.

(34) Zaitsev, A. M. *Optical Properties of Diamond*; Springer Berlin Heidelberg: Berlin, Heidelberg, 2001. https://doi.org/10.1007/978-3-662-04548-0.

(35) Mutti, P.; Bottani, C. E.; Ghislotti, G.; Beghi, M.; Briggs, G. A. D.; Sandercock, J. R. Surface Brillouin Scattering—Extending Surface Wave Measurements to 20 GHz. In *Advances in Acoustic Microscopy*; Briggs, A., Ed.; Advances in Acoustic Microscopy; Springer: Boston, MA, 1995; Vol. 1, pp 249–300. https://doi.org/10.1007/978-1-4615-1873-0_7.

(36) Kargar, F.; Penilla, E. H.; Aytan, E.; Lewis, J. S.; Garay, J. E.; Balandin, A. A. Acoustic Phonon Dispersion Engineering in Bulk Crystals via Incorporation of Dopant Atoms. *Applied Physics Letters* **2018**, *112* (19), 191902. https://doi.org/10.1063/1.5030558.

(37) Brunet, F.; Germi, P.; Pernet, M.; Deneuville, A.; Gheeraert, E.; Laugier, F.; Burdin, M.; Rolland, G. The Effect of Boron Doping on the Lattice Parameter of Homoepitaxial Diamond Films. *Diamond and Related Materials* **1998**, *7* (6), 869–873. https://doi.org/10.1016/s0925-9635(97)00316-6.

(38) Liu, X.; Chang, Y. Y.; Tkachev, S. N.; Bina, C. R.; Jacobsen, S. D. Elastic and Mechanical Softening in Boron-Doped Diamond. *Scientific Reports* **2017**, *7* (1), 42921. https://doi.org/10.1038/srep42921.

(39) Wang, X.; Shen, X.; Sun, F.; Shen, B. Influence of Boron Doping Level on the Basic Mechanical Properties and Erosion Behavior of Boron-Doped Micro-Crystalline Diamond (BDMCD) Film. *Diamond and Related Materials* **2017**, *73*, 218–231. https://doi.org/10.1016/J.DIAMOND.2016.09.025.

(40) Giannozzi, P.; Baroni, S.; Bonini, N.; Calandra, M.; Car, R.; Cavazzoni, C.; Ceresoli, D.; Chiarotti, G. L.; Cococcioni, M.; Dabo, I.; Dal Corso, A.; de Gironcoli, S.; Fabris, S.; Fratesi, G.; Gebauer, R.; Gerstmann, U.; Gougoussis, C.; Kokalj, A.; Lazzeri, M.; Martin-Samos, L.; Marzari, N.; Mauri, F.; Mazzarello, R.; Paolini, S.; Pasquarello, A.; Paulatto, L.; Sbraccia, C.; Scandolo, S.; Sclauzero, G.; Seitsonen, A. P.; Smogunov, A.; Umari, P.; Wentzcovitch, R. M. QUANTUM ESPRESSO: A Modular and Open-Source Software Project for Quantum of Materials. *Journal of Physics: Condensed Matter* **2009**, *21* (39), 395502. https://doi.org/10.1088/0953-8984/21/39/395502.

(41) Warren, J. L.; Yarnell, J. L.; Dolling, G.; Cowley, R. A. Lattice Dynamics of Diamond. *Physical Review* **1967**, *158* (3), 805–808. https://doi.org/https://doi.org/10.1103/PhysRev.158.805.

(42) Zheng, F.; Zhang, P. Phonon Unfolding: A Program for Unfolding Phonon Dispersions of Materials. *Computer Physics Communications* **2017**, *210*, 139–144. https://doi.org/10.1016/J.CPC.2016.09.005.

(43) Klemens, P. G. Thermal Conductivity and Lattice Vibrational Modes; Seitz, F., Turnbull, D., Eds.; Solid State Physics; Academic Press, 1958; Vol. 7, pp 1–98. https://doi.org/http://dx.doi.org/10.1016/S0081-1947(08)60551-2.

(44) Han, Y. J.; Klemens, P. G. Anharmonic Thermal Resistivity of Dielectric Crystals at Low Temperatures. *Physical Review B* **1993**, *48* (9), 6033–6042. https://doi.org/10.1103/PhysRevB.48.6033.

(45) Callaway, J. Model for Lattice Thermal Conductivity at Low Temperatures. *Physical Review* **1959**, *113* (4), 1046. https://doi.org/10.1103/PhysRev.113.1046.

(46) Balandin, A.; Wang, K. L. Significant Decrease of the Lattice Thermal Conductivity Due to Phonon Confinement in a Free-Standing Semiconductor Quantum Well. *Physical Review B* **1998**, *58* (3), 1544. https://doi.org/10.1103/PhysRevB.58.1544.

(47) Liu, W. L.; Shamsa, M.; Calizo, I.; Balandin, A. A.; Ralchenko, V.; Popovich, A.; Saveliev, A. Thermal Conduction in Nanocrystalline Diamond Films: Effects of the Grain Boundary Scattering and Nitrogen Doping. *Applied Physics Letters* **2006**, *89* (17), 171915. https://doi.org/10.1063/1.2364130.

(48) Tarelkin, S.; Bormashov, V.; Kuznetsov, M.; Buga, S.; Terentiev, S.; Prikhodko, D.; Golovanov, A.; Blank, V. Heat Capacity of Bulk Boron-Doped Single-Crystal HPHT Diamonds in the Temperature Range from 2 to 400 K. *Journal of Superhard Materials* **2016**, *38* (6), 412–416. https://doi.org/10.3103/S1063457616060058.

(49) Zou, J.; Kotchetkov, D.; Balandin, A. A.; Florescu, D. I.; Pollak, F. H. Thermal Conductivity of GaN Films: Effects of Impurities and Dislocations. *Journal of Applied Physics* **2002**, *92* (5), 2534–2539. https://doi.org/10.1063/1.1497704.

(50) Klemens, P. G. Heat Conduction in Solids by Phonons. *Thermochimica Acta* **1993**, *218*, 247–255. https://doi.org/10.1016/0040-6031(93)80426-B.

(51) Liu, W.; Balandin, A. A. Thermal Conduction in AlxGa1-XN Alloys and Thin Films. *Journal of Applied Physics* **2005**, *97* (7), 073710. https://doi.org/10.1063/1.1868876.

(52) Herring, C. Role of Low-Energy Phonons in Thermal Conduction. *Physical Review* **1954**, *95* (4), 954–965. https://doi.org/10.1103/PhysRev.95.954.

(53) Slack, G. A. Effect of Isotopes on Low-Temperature Thermal Conductivity. *Physical Review* **1957**, *105* (3), 829–831. https://doi.org/10.1103/PhysRev.105.829.

(54) Kotchetkov, D.; Zou, J.; Balandin, A. A.; Florescu, D. I.; Pollak, F. H. Effect of Dislocations on Thermal Conductivity of GaN Layers. *Applied Physics Letters* **2001**, *79* (26), 4316–4318. https://doi.org/10.1063/1.1427153.

(55) Sparavigna, A. Influence of Isotope Scattering on the Thermal Conductivity of Diamond. *Physical Review B* **2002**, *65* (6), 064305. https://doi.org/10.1103/PhysRevB.65.064305.

(56) Liu, M. S.; Bursill, L. A.; Prawer, S.; Beserman, R. Temperature Dependence of the First-Order Raman Phonon Line of Diamond. *Physical Review B* **2000**, *61* (5), 3391–3395. https://doi.org/10.1103/PhysRevB.61.3391.

(57) Solin, S. A.; Ramnas, A. K. Raman Spectrum of Diamond. *Physical Review B* **1970**, *1* (4), 1687–1698. https://doi.org/https://doi.org/10.1103/PhysRevB.1.1687.

(58) Lee, K. C.; Sussman, B. J.; Nunn, J.; Lorenz, V. O.; Reim, K.; Jaksch, D.; Walmsley, I. A.; Spizzirri, P.; Prawer, S. Comparing Phonon Dephasing Lifetimes in Diamond Using Transient Coherent Ultrafast Phonon Spectroscopy. *Diamond and Related Materials* **2010**, *19* (10), 1289–1295. https://doi.org/10.1016/J.DIAMOND.2010.06.002.

(59) Perdew, J. P.; Burke, K.; Ernzerhof, M. Generalized Gradient Approximation Made Simple. *Physical Review Letters* **1996**, *77* (18), 3865–3868. https://doi.org/10.1103/PhysRevLett.77.3865.

(60) Baroni, S.; de Gironcoli, S.; Dal Corso, A.; Giannozzi, P. Phonons and Related Crystal Properties from Density-Functional Perturbation Theory. *Reviews of Modern Physics* **2001**, *73* (2), 515–562. https://doi.org/10.1103/RevModPhys.73.515.